\documentclass[a4paper, 12pt]{article}
\usepackage[english]{babel}
\usepackage[a4paper, inner=2cm, outer=2cm, top=3cm, bottom=3cm]{geometry}
\usepackage[dvipsnames]{xcolor}
\usepackage{mathtools}
\usepackage{pstricks}
\usepackage{caption}
\usepackage{graphicx}
\usepackage{amsmath}
\usepackage{amssymb}
\usepackage{mathcomp}
\usepackage{textcomp}
\usepackage{enumitem}
\usepackage{physics}
\usepackage{romannum}
\usepackage[doublespacing]{setspace}
\usepackage{subcaption}
\usepackage{hyperref}

\hypersetup{
  colorlinks   = true,
  urlcolor     = Blue, 
  linkcolor    = Black,
  citecolor    = Black
}

\renewcommand{\thesection}{\Roman{section}} 

\title{Traversability of quantum improved wormhole solution}
\author{R. Moti and A. Shojai\\
\textit{\small Department of Physics, University of Tehran, Tehran, Iran}}
\date{}

\begin{document}
\maketitle
\pagenumbering{arabic}
\begin{abstract}
We have investigated the problem of traversability of wormholes in the framework of quantum improvement of gravity theory arising from functional renormalization group methods to describe the asymptotic safe quantum gravity. We have shown that improved pseudospherical wormholes could be traversable with non--exotic matter, while spherical ones could not. This is done within a specific model of improvement.
\end{abstract}

\section{Introduction}
The concept of wormholes and the possibility that they may be traversable is an interesting issue in general relativity. For Einstein theory of gravity, it can be shown\cite{Morris} that it is impossible to have a traversable wormhole with ordinary matter, and exotic one is needed.
Extension of the concept to other gravitational theories, like some classical modified theories (e.g. Einstein--Gauss--Bonett gravity\cite{EGB-G}, Lovelock gravity\cite{L-G}, Einstein--Born--Infeld gravity\cite{EBI-G}, and others\cite{Oth-ext})  and some quantum corrected ones\cite{Q-C} can be found in the literature.

The possibility of a traversable wormhole solution is usually searched for by considering a static spherical symmetric solution of the form
\begin{equation}
    \dd{s}^2 = e^{2\Phi(r)} \dd{t}^2 - \frac{\dd{r}^2}{1-b(r)/r} -r^2\dd{\Omega}_{2(s/p)}
\end{equation}
where the $\dd{\Omega}_{2(s/p)}$ is the 2--sphere or 2--pseudosphere line element.

It is known that the wormhole solution could be possible, provided that at least the following three conditions are fufilled by the redshift function $e^{2\Phi(r)}$ and the shape function $b(r)$ \cite{Morris, Cataldo}:
\begin{itemize}
  \item[\Romannum{1})] No event horizon near the throat.
  \item[\Romannum{2})] Existence of the throat, determined by the embedding space of the radial coordinate $r$, such that it decreases from $\infty$ to a minimum value $b(r_t)$ at the throat and then increases again, to $\infty$  after the throat.
This means $b(r_t)=r_t$, where $r_t$ is the throat location.
  \item[\Romannum{3})] The flare--out condition, which reads as the condition that the embedded hypersurface $z=z(r)$ satisfies $\eval{\dv[2]{r}{z}}_{r=r_{t}} > 0$, leading to  $b'(r_t) < 1$, where  $'$ denotes differentiation with respect to $r$.
\end{itemize}
Then, the traversability of the throat depends on the existence of a flow that couples to the geometry in such a way that the null energy condition is satisfied.

There are two important questions that should be addressed. They are: ``{\em how the quantum effects can change the traversability of wormholes?} '', and ``{\em is it possible to have a quantum corrected wormhole that is traversable with non--exotic matter?} ''.
To answer these questions one has to have a quantum theory of gravity. Unfortunately there is no established one, yet. 

One way to include the quantum effects is to consider gravity as an asymptotic safe effective quantum field theory, which runs to a non--Gausssian fixed point at UV\cite{as}. In this condition, the theory becomes UV complete.
Evidence for the existence of such a fixed point for the gravity renormalization group flow, is confirmed by various methods (see references in \cite{Weinberg-inf}).
The truncated renormalization group method is one of the successful methods in this way\cite{Reuter-1st}. Unfortunately, solving the exact renormalization group equation to derive the effective average action is complex, if not impossible. Therefore the effects of this quantization method are usually considered as a correction to the classical theory and studied through an effective theory obtained by improving the classical coupling constant to the running one which is derived from the solution for the $\beta$-function\cite{Reuter-1st,Reuter-2nd,Reuter-3rd}.

This can be done with different strategies as it is discussed in\cite{our3}. The most physical way of quantum improvement of the field equations, is the action improvement presented in\cite{our3}. In this method the action functional is covariantly improved leading to the modified field equations
\begin{equation} 
G_{\mu\nu} = 8\pi G(\chi) T_{\mu\nu} + G(\chi) X_{\mu\nu} (\chi) \ ,\label{IEM}
\end{equation}
where $G(\chi)$ is the improved coupling constant which is a function of curvature invariants $\chi$, and the covariant tensor $X_{\mu\nu}$ is defined as
\[
    X_{\mu\nu}(\chi) = \Big( \nabla_{\mu}\nabla_{\nu} - g_{\mu\nu}\square \Big) G(\chi)^{-1}   - \frac{1}{2} \bigg( R\mathcal{K}(\chi) \frac{\delta\chi}{\delta g^{\mu\nu}} +
 \]
\begin{equation}
 \partial_{\kappa}\Big (R\mathcal{K}(\chi)\frac{\partial\chi}{\partial (\partial_{\kappa}g^{\mu\nu})}\Big) + \partial_{\kappa}\partial_{\lambda}\Big (R\mathcal{K}(\chi)\frac{\partial\chi}{\partial (\partial_{\lambda}\partial_{\kappa}g^{\mu\nu})}\Big ) \bigg ) \ ,
\end{equation}
with $\mathcal{K}(\chi)\equiv  \frac{2\pdv*{G(\chi)}{\chi}}{G(\chi)^2}$ \cite{our4}.
  
Although the exact renormalization group equation determines the dynamics of this prescribed field $G(\chi)$, the functional renormalization group methods beside other assumptions \cite{Reuter-1st, Souma}, leads to the antiscreening running gravitational coupling as

\begin{equation}
  G(\chi) = \frac{G_0}{1+f(\chi)}
\end{equation}
where $f(\chi) \equiv \xi /\chi$, and $\chi$ is a function of curvature invariants of dimension length square.
This is called antiscreening running coupling as it goes to zero at small length scale ($\chi\rightarrow 0$).

The small scaling constant $\xi$ can be written as $G_0 \omega_q \xi_0 $ with the condition that the reference constant $ G_0 $ be the experimentally observed value of Newton's constant $G_N$, and $\xi_0$ is a constant of order of one and $\omega_q = \frac{4}{\pi}(1-\frac{\pi^{2}}{144})$ \cite{Bonanno & Reuter}.

The quantum correction term $X_{\mu\nu}$ depends on the scaling factor and up to the first order, we would have\cite{our3, our4}
\begin{equation}
  X_{\mu\nu} \simeq \nabla_{\mu}\nabla_{\nu} G(\chi)^{-1} - g_{\mu\nu}\square G(\chi)^{-1} \ .
\end{equation}

The renormalization group parameter $\chi$ is the scaling parameter of the theory. As it is discussed in detail in \cite{our3}, it can be argued that  the best scaling parameters of the spacetime are the components of the Riemann tensor, which physically describe tidal forces. To save the general covariance of the action after improvement, they should contribute in $\chi$ in terms of curvature invariants. Thus, in general, $\chi$ could be a well--defined function of all independent curvature invariants such as $R, R_{\alpha\beta}R^{\alpha\beta}, R_{\alpha\beta\gamma\delta}R^{\alpha\beta\gamma\delta},\cdots$.

Unfortunately, one of the unwanted features of this method is that there is no unique way to fix the form of $\chi$, in general yet\cite{Pawlowski}. But for non--vacuum solutions, considering various conditions such as the behavior of singularities and energy conditions\cite{our4}, one can restrict the possible choices. Simple choices for non--vacuum solutions are $\chi=R^{-1}$ or 
$\chi=(R_{\alpha\beta}R^{\alpha\beta})^{1/2}$.

Here we are looking for the effects on wormhole traversability within the context of the above mentioned quantum improved theory.
In the next sections, such solutions are searched for, in the improved spacetime, assuming $\chi=R^{-1}$ as an example.

Such a choice is simple, makes the comparision with $f(R)$ solutions\cite{Hindmarsh} possible, and for cosmological solutions coincides with  the Hubble parameter (which is a natural scale)\cite{Reuter-3rd}.

Effects of other choices for $\chi$ on the field equations are studied in the literature\cite{our3, our2, Pawlowski}.

We will study the possibility of having traversable wormholes with non--exotic matter for the quantum improved equations. In the Einstein theory of gravity, the throat condition would not be satisfied unless the radial pressure, $p_r$, of the matter content be such that the dimensionless exoticity function $\xi_e=-(p_{r}+\rho)/\rho$ is positive, where $\rho$ is the energy density\cite{Morris}. The satisfaction of this condition is nothing but the violation of the energy conditions, and is possible by considering some \textit{exotic matter}\cite{Lobo-Book}. This situation could be described best by investigating the behavior of a null geodesic congruence, passing through the wormhole throat. The expansion rate of the cross section of this congruence should increase after a period of decreasing before passing through the throat. This sign change of the expansion rate at the throat could happen only by a repulsive behavior of gravity, which in the context of the Raychaudhuri equation leads to the violation of the energy conditions\cite{Poisson}. We would show that the quantum corrections change this description. The repulsion can be caused by the quantum corrections to the geometry. Indeed, as discussed in\cite{our4}, the attractiveness behavior of gravity depends not only on the energy conditions, but also on the quantum effects.

\section{Pseudospherical traversable solution}
Let us first consider pseudospherical wormholes. A general static hyperbolic solution is of the form
\begin{equation}
    \dd{s}^2 = e^{2\Phi(r)} \dd{t}^2 - \frac{\dd{r}^2}{1-b(r)/r} -r^2\dd{\Omega}_{2(p)} \label{PSM}
\end{equation}
where as it is noted earlier $e^{2\Phi}$ is the redshift function (and thus $\Phi$ is the tidal potential) and $b$ is called the shape function.

On using the improved equations of motion (\ref{IEM}), with an anisotropic fluid 
$$T^{\mu}_{\nu} = \text{Diag} [\rho,-p_r,-p_l,-p_l]$$ 
one arrives at the following field equations
 \begin{align}
    8\pi G_0 \rho  = &\bigl(1+f\bigr) \frac{b^{'}-2}{r^2} -(1-\frac{b}{r}) (f''+\frac{2}{r} f')+ \frac{b^{'}r-b}{2r^2}f'  \label{IEQ-tt-p}\\
    8\pi G_0 p_r  = & -\bigl(1+f\bigr) \left(\frac{b}{r^3} -\frac{2}{r^2} -\frac{2\Phi^{'}}{r}(1-\frac{b}{r}) \right) + (1-\frac{b}{r}) \left( \Phi^{'}+\frac{2}{r}\right) f'   \label{IEQ-rr-p} \\
    8\pi G_0 p_l  = & -\bigl(1+f\bigr) \left( \frac{b'r-b}{2r^2} (\Phi'+\frac{1}{r} ) - (1-\frac{b}{r})(\Phi''+\Phi'^2+\frac{\Phi'}{r}) \right)  \nonumber\\
    & + \left((1-\frac{b}{r}) ( \Phi^{'}+\frac{1}{r})- \frac{b^{'}r-b}{2r^2}\right) f' + (1-\frac{b}{r})f'' \ . \label{IEQ-pp-p}
  \end{align}

Although all the independent curvature invariants could practically be considered in defining the cutoff function, but here for simplicity we just choose the scalar curvature, i.e. we set $f =\xi R$. We shall comment on other choices in the concluding remarks.

In addition, to have a specific proper model we choose anistropic matter with linear state equation, $p_r = \omega \rho$. Putting everything in place, equations (\ref{IEQ-tt-p}) and (\ref{IEQ-rr-p}) can be combined to give the following relation between the tidal potential and the shape function 
\begin{equation}
     \omega b' r +b -2(\omega+1)r  -2 \Phi' r^2 (1-\frac{b}{r})= \frac{r^3}{1+f} \left((1-\frac{b}{r})(\omega f''+\frac{2}{r} (1+\omega)f'+\Phi'f')-\omega\frac{b^{'}r-b}{2r^2}f'  \right) \ .
\end{equation}

Usually there is a special attention to the zero--tidal--force models, i.e. $\Phi=0$. It can be shown that tidal force seen by a local stationary observer who travels through the wormhole, should not exceed the earth gravity\cite{Morris}. This condition besides the travel speed considerations lead to zero tidal force models. Therefore, to compare the results of our model with the classical ones, we put $\Phi=0$ in the remaining calculations.

For a zero--tidal--force spacetime, the dimensionless shape function $\tilde{b} \equiv b/r_t$ is the solution of
\begin{equation}
  \omega \dot{\tilde{b}} u +\tilde{b} -2(\omega+1)u = \frac{u^3}{1+f} \left((1-\frac{\tilde{b}}{u})(\omega \ddot{f}+\frac{2}{u} (1+\omega)\dot{f})-\omega\frac{\dot{\tilde{b}}u-\tilde{b}}{2u^2}\dot{f}  \right) \label{be-p}
\end{equation}
where a dot over any quantity, denotes derivative with respect to the dimensionless variable $u \equiv r/r_t$. 

Since $f$ is proportional to $\xi$, the right hand side of this dimensionless equation is just a small correction, and thus we can solve it by  iteration. In the zeroth order the right hand side should be ignored and with the initial condition $\tilde{b}_{(0)}(1) = 1$, the solution is $\tilde{b}_{(0)}(u) =2u-u^{-1/\omega}$.
To obtain the first order solution, we note that only the solution near the throat is needed. Expanding around  $u=1$ and using the zeroth order solution, the equation (\ref{be-p}) in the first order of iteration is 
\begin{multline}
   \omega \dot{\tilde{b}}_{(1)}u + \tilde{b}_{(1)}-2(\omega+1) u = \\
   \zeta \frac{1+3\omega}{\omega^2( \omega -2\zeta)^2} \Bigl( \omega^2(1+\omega)( 7u-8) + 2\zeta(\omega(1+\omega) +(u-1)( 1-3\omega-4\omega^2))\Bigr)
\end{multline}
where $\zeta \equiv \xi/r_t^2$. 

As a result, the shape function becomes
\begin{multline}
  \tilde{b}(u) \simeq  \frac{1}{\omega^2(\omega-2\zeta)^2} \times \\
  \Bigl(\omega^4(2u-u^{-1/\omega}) -\zeta\omega^2(8+32\omega+24\omega^2-u(7+13\omega)-u^{-1/\omega}(1+15\omega+24\omega^2)) \\
  -2\zeta^2  (1-\omega-17\omega^2-15\omega^3 +\omega^2(7+15\omega) u^{-1/\omega} +u(-1+\omega+8\omega^2)) \Bigr) \label{b-p}
\end{multline}
around the throat. 
$\Phi=0$ and the above shape function present a solution of the improved Einstein's equations around the throat.

The validity and trversability of this solution can be guaranteed by implying the following four conditions which restrict the state equation parameter $\omega$ and the quantum correction parameter $\zeta$:

\begin{itemize}
  \item Throat Condition: As stated before, this condition is in fact three conditions: No event horizon condition, the $b(r_t)=r_t$  condition and the flare--out condition.
  
  Clearly there is no event horizon for zero--tidal--force model. The second condition was already applied as $\tilde{b}(1)=1$. Finally, the third one leads to the relation $\dot{\tilde{b}}(1)<1$. For our model, using equation (\ref{b-p}), this is nothing except a restriction on the parameters $\omega$ and $\zeta$:
  \begin{equation}
    \frac{(1+2\omega)\omega^2-\zeta(1+6\omega+7\omega^2)}{\omega^2(\omega-2\zeta)} <1 \ .
  \end{equation}
  The region of validity of this condition is shown by dashed lines in the figure (\ref{fig1}).

\begin{figure}
    \centering
		\includegraphics[width=0.55\textwidth]{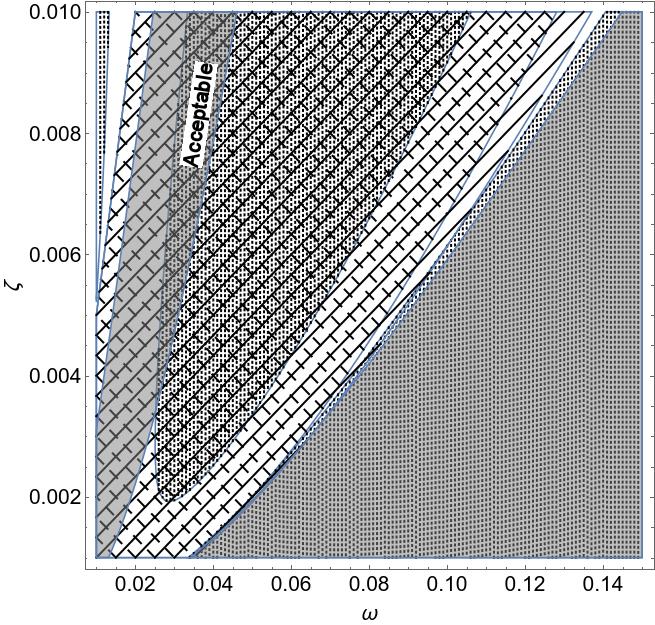}
	\caption{The region of validity of the traversability conditions for pseudospherical solution (at $u=1.008$). Dashed, dotted and thick lines represent the validity region of throat, radial null energy condition and antiscreening conditions, respectively. Grey area denotes the region of validity transverse null energy condition. There is a big area where all the conditions are satisfied and the wormhole is traversable, with normal matter.}
	\label{fig1}
\end{figure} 

    \item Antiscreening Condition: 
As it is discussed in the introduction section, the functional renormalization group method leads to an antiscreening running coupling. To be in the region of validity of the method, we have to have the criteria of antiscreening running gravitational coupling, i.e.
  \begin{equation}
    \eval{f(\chi)}_{u\rightarrow 1} = \frac{2\zeta(-\omega^2+\zeta(1+6\omega+3\omega^2)}{\omega^2(\omega-2\zeta)}>0 \ . \label{Index Inertia S}
  \end{equation}
  This restricts the parameters $(\omega,\zeta)$, to the region denoted by thick lines in figure (\ref{fig1}).
  
  \item Null Energy Condition: 
The radial null energy condition for the radial pressure (i.e. $\rho+p_r>0$) at the throat reads as

\begin{equation} 
   8\pi G_N(\rho+p_r) = \eval{\left((1+ f) \left( \frac{\dot{\tilde{b}}}{u^2} - \frac{\tilde{b}}{u^3} \right)+ \frac{\dot{\tilde{b}}u-\tilde{b}}{2u^2}\dot{f}\right) }_{u=1} > 0 \ . 
\end{equation}
This sets a constraint on the cutoff identification function, $f$. On using the second and third parts of the throat condition, we arrive at the following relation
\begin{equation}
  \eval{\left(\frac{1+ f}{u} + \frac{\dot{f}}{2}\right) }_{u=1} < 0 \ .
\end{equation}
  
Using  \eqref{IEQ-tt-p}, for zero--tidal--force case, the radial null energy condition could be rewritten as
  \begin{equation}
   \eval{(1+\omega) \left( (1+f)(\dot{\tilde{b}}-2)+\frac{\dot{\tilde{b}} u -\tilde{b}}{2} \dot{f} \right)}_{u=1}  > 0
  \end{equation}
and for the solution found for the improved Einstein's equations, this gives the following constraint 
  \begin{equation}
 (\omega^2-\zeta(1+6\omega+3\omega^2)) \left(\omega^3-\zeta(5\omega^2+\omega^3)+\zeta^2(3+18\omega+11\omega^2)\right) > 0 \ .
  \end{equation}
The region of validity of this constraint is plotted by dotted lines in  figure (\ref{fig1}). 
It has to be noted that since we are dealing with state equation $p_r=\omega\rho$, the radial null energy condition simplifies to $8\pi G_0(1+\omega)\rho>0$ and thus positive $\omega$ is the sign that we have non--exotic matter.
 
 On the other hand, the null energy condition on the transverse parts of stress tensor (i.e. $\rho+p_l>0$), can be checked by using equations (\ref{IEQ-tt-p}) and (\ref{IEQ-pp-p}). This gives additional restrictions on the parameters and is denoted by grey area in figure (\ref{fig1}). 
  
  It is remarkable that the regions of validity of these three conditions have a notable overlap which contains positive values of $(\omega,\zeta)$.
   This endorse that there could be traversable wormhole solution with non--exotic matter.
Also note that the way we adopted the conditions, means that the weak energy conditions are also satisfied.  

  \item Index Inertia: Finally there is one more condition to consider. To save the signature of the metric it is necessary that $g_{rr} < 0$. Since we are interested in the metric behavior near the throat, the index inertia condition leads to
  \begin{equation}
    \eval{ \tilde{b}(u) < u }_{u\rightarrow 1^{+}}.  
  \end{equation}
By using equations \eqref{b-p} and \eqref{Index Inertia S} we find this fourth condition on the $(\omega,\zeta)$ as:
   \begin{multline}
   \frac{1}{\omega^2(\omega-2\zeta)^2} \times \\
  \Bigl(\omega^4(2u-u^{-1/\omega}) -\zeta\omega^2(8+32\omega+24\omega^2-u(7+13\omega)-u^{-1/\omega}(1+15\omega+24\omega^2)) \\
  -2\zeta^2  (1-\omega-17\omega^2-15\omega^3 +\omega^2(7+15\omega) u^{-1/\omega} +u(-1+\omega+8\omega^2)) \Bigr)  < u \ .
  \end{multline}
 As it can be seen from figure (\ref{fig2}), although the classical solution changes sign at some $u>1$ (and thus violates the signature condition of metric), the quantum improved one shows the correct signature for the radial component of the metric near the throat. This is true for the acceptable values of $(\omega,\zeta)$ obtained from previous conditions and is illustrated for some special value of the $(\omega,\zeta)$ in figure (\ref{fig2}).
      
\begin{figure}
    \centering
		\includegraphics[scale=0.4]{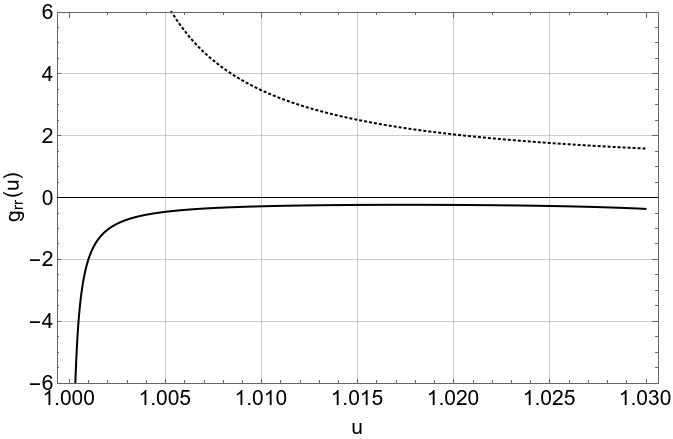}
	\caption{The quantum improved radial component of the metric, $g_{rr}$ (for pseudospherical solution) plotted by a thick line for some specific $(\omega,\zeta) = (0.030,0.007)$. The dashed line is the corresponding one for the classical solution  which explicitly violates the signature of the metric.}
	\label{fig2}
\end{figure}
\end{itemize}
As we can see, it is quite possible to find a quantum improved traversable pseudospherical solution, for a set of $(\omega,\zeta)$ parameters that saves the causal structure, satisfies  null energy condition, and the matter is non--exotic.

\begin{figure}
    \centering
		\includegraphics[scale=0.4]{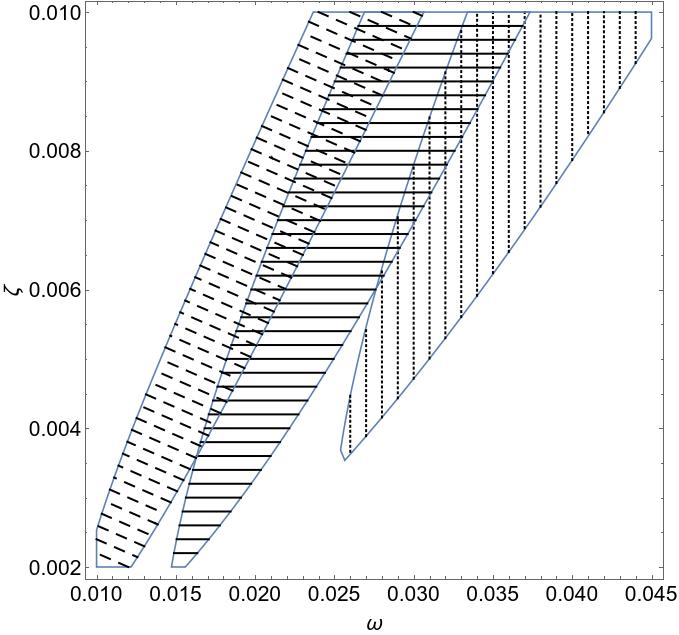}
	\caption{The acceptable region of parameters $(\omega,\zeta)$ for different values of $u$ (for the pseudospherical solution). Dotted lines for $u=1.008$, thick lines for $u=1.005$, and dashed lines for $u=1.003$.}
	\label{fig1a}
\end{figure}

It has to be noted here that the acceptable region of parameters $(\omega,\zeta)$ slightly depend on the location of passenger near the throat. This can be seen in figure (\ref{fig1a}). As we can see to make a pseudospherical wormhole traversable, one needs to have matter with a slightly changing value of $\omega$ near the throat.

It has to be noted that such a traversable wormhole is not at the astrophysical scales. It is a quantum wormhole, applicable in the very early universe and for travelers having lengths of planck scale. This is discussed in the Appendix.

\section{Spherical traversable solution}
There is another possibility for which the space is spherical, instead of pseudospherical. Looking for such a wormhole solution, one should set
\begin{equation}
   \dd{s}^2 = e^{2\Phi(r)} \dd{t}^2 - \frac{\dd{r}^2}{1-b(r)/r} -r^2\dd{\Omega}_{2(s)} \ . \label{SSM}
\end{equation}
For an anistropic fluid  $ T^{\mu}_{\nu} = \text{Diag} [\rho(r),-p_r(r),-p_l(r),-p_l(r)] \ $, the improved field equations \eqref{IEM} lead to the
\begin{align}
    8\pi G_0 \rho  = &(1+f) \frac{b^{'}}{r^2} -  (1-\frac{b}{r}) (f''+\frac{2}{r} f')+ \frac{b^{'}r-b}{2r^2}f' \ , \label{IEQ-tt-s}\\
    8\pi G_0 p_r  = & -(1+f) \left(\frac{b}{r^3} -\frac{2\Phi^{'}}{r}(1-\frac{b}{r}) \right) + (1-\frac{b}{r}) \left( \Phi^{'}+\frac{2}{r}\right) f'  \ ,  \label{IEQ-rr-s} \\
    8\pi G_0 p_l  = & -(1+f) \left( \frac{b'r-b}{2r^2} (\Phi'+\frac{1}{r} ) - (1-\frac{b}{r})(\Phi''+\Phi'^2+\frac{\Phi'}{r}) \right) \nonumber\\
    & + (1-\frac{b}{r}) \left(( \Phi^{'}+\frac{1}{r})f'+f''\right)  - \frac{b^{'}r-b}{2r^2}f' \ . \label{IEQ-ll-s}
\end{align}
As in the previous section, we choose the simple model $f=\xi R$.
For a fluid with linear state equation $p_r=\omega\rho$, the proper shape function is the solution of
\begin{equation}
     \omega b' r +b -2 \Phi' r^2 (1-\frac{b}{r})= \frac{r^3}{1+f} \left((1-\frac{b}{r})(\omega f''+\frac{2}{r} (1+\omega)f'+\Phi'f')-\omega\frac{b^{'}r-b}{2r^2}f'  \right)
\end{equation}
and with the zero--tidal--force assumption, the dimensionless shape function $\tilde{b} \equiv b/r_t$ should be the solution of this equation
\begin{equation}
  \omega \dot{\tilde{b}} u +\tilde{b} = \frac{u^3}{1+f} \left((1-\frac{\tilde{b}}{u})(\omega \ddot{f}+\frac{2}{u} (1+\omega)\dot{f})-\omega\frac{\dot{\tilde{b}}u-\tilde{b}}{2u^2}\dot{f}  \right) \ . \label{be-s}
\end{equation}
We can solve this equation by iteration as it is done for the pseudospherical case.
By considering the initial condition $\tilde{b}_{(0)}(1) = 1$, the zero order solution becomes $\tilde{b}_{(0)}(u) = u^{-1/\omega}$, and in the first order of iteration, equation \eqref{be-s} can be written up to first order near the throat as
\begin{equation}
  \omega \dot{\tilde{b}}_{(1)} u +\tilde{b}_{(1)} = \xi(1+\omega) (1+3\omega)  \frac{r_t^2 \omega^2 (-8+7u) +2\xi(1-5\omega+u(-1+4\omega)) }{\omega^2(2\xi+ r_t^2 \omega)^2} 
\end{equation}
leading to the following shape function
\begin{multline}
\tilde{b}(u) \simeq  \frac{1}{\omega^2 (2\zeta+\omega)^2} \times \\
 \Bigl( \omega^4u^{-1/\omega} +\zeta \bigl[ 7  \omega^2(1+3\omega)u^{1-1/\omega}  - 8  \omega^2 (1+4\omega+3\omega^2) + \omega^2 u^{-1/\omega}(1+15\omega+24\omega^2) \bigr] \\
  2 \zeta^2 \bigl[ u^{-1/\omega} \omega^2 (7+15\omega) + u (-1+\omega+12\omega^2)+(1-\omega-17\omega^2-15\omega^3)   \bigr]\Bigr) \ , \label{b-s}
\end{multline}
where again $\zeta \equiv \xi/r_t^2$.

Now we can check the traversability conditions for this wormhole solution and find the allowed parameters $(\omega,\zeta)$:
\begin{itemize}
  \item Throat Condition: As stated before, only the flare--out condition should be investigated since the other two are satisfied by our solution. It leads  to the $\dot{\tilde{b}}(u=1)<1$, and thus the acceptable parameters are the ones which satisfy the condition
  \begin{equation}
\frac{(1+2\omega)\omega^2-\zeta(1+6\omega+7\omega^2)}{\omega^2(\omega-2\zeta)} <1 \ .\label{C-s-1}
  \end{equation}
In figure (\ref{fig3}) this region is shown by dashed lines.
      
\begin{figure}
    \centering
		\includegraphics[width=0.50\textwidth]{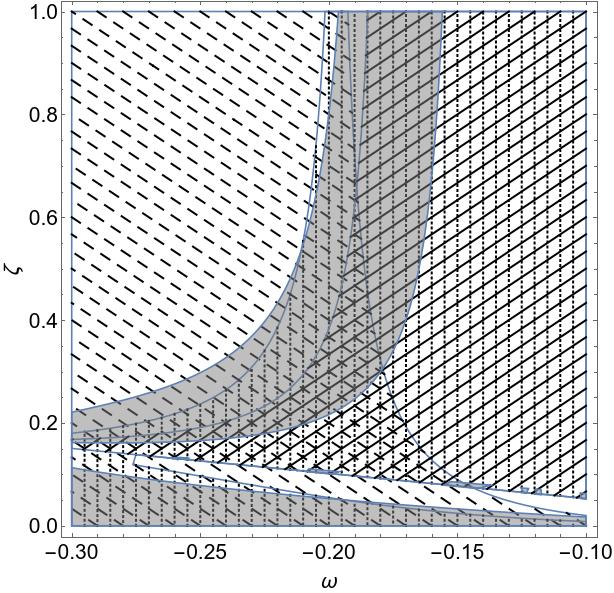}
	\caption{The region of validity of the traversability conditions, for spherical solution (at $u=1.008$). Dashed, dotted and thick lines represent the validity region of throat, radial null energy condition and antiscreening conditions, respectively. Grey area denotes the region of validity transverse null energy condition. There is a big area where all the conditions are satisfied and the wormhole is traversable, with exotic matter.}
	\label{fig3}
\end{figure} 

  \item Antiscreening Condition: In parallel with the previous section, we have to set $f>0$.
For our model this conditions leads to
\begin{equation}
\frac{2\zeta(\zeta+6\omega\zeta+\omega^2(1+3\zeta))}{\omega^2(\omega+2\zeta)} > 0 \ .
\end{equation}

 In figure (\ref{fig3}) the allowed region is denoted by thick lines.

  \item Null Energy Condition: The radial null energy condition is again $8\pi G_0 (1+\omega) \rho>0$. Having $\rho$ derived from \eqref{IEQ-tt-s}, and the allowed values of $(\omega,\zeta)$ from \eqref{C-s-1} at the throat, we get 
  \begin{equation}
   \eval{(1+\omega) \left( (1+f)(\dot{\tilde{b}}-2)+\frac{\dot{\tilde{b}} u -\tilde{b}}{2} \dot{f} \right)}_{u=1}  > 0 \ ,
  \end{equation}
  which for $f=\xi R$ model becomes
  \begin{multline}
    -(\omega^6+2\omega^7) - \zeta (8\omega^5+16\omega^6) +\zeta^2(2\omega^2+20\omega^3+36\omega^4+12\omega^5+18\omega^6) \\
    +\zeta^3 (1+22\omega+150\omega^2+364\omega^3+329\omega^4+126\omega^5)+\zeta^4(14+128\omega+356\omega^2+352\omega^3+126\omega^4) > 0 \ .
  \end{multline}
  Regions with dotted lines in figure (\ref{fig3}), shows the area of validity of this condition.

Also the region of validity of transverse null energy condition is shown by grey area in figure (\ref{fig3}).
 
 As one can see, the acceptable region of parameters $(\omega,\zeta)$ indicates that we are dealing with exotic matter, as $\omega$ is now negative.
  \item Index Inertia: The signature is saved $(g_{rr} < 0)$ if the condition
  \begin{equation}
    \eval{ \tilde{b}(u) < u }_{u\rightarrow 1^{+}}  
  \end{equation}
   is satisfied. By equation \eqref{b-s} this becomes:
   \begin{multline}
   \frac{1}{\omega^2 (2\zeta+\omega)^2} \times \\
 \Bigl( \omega^4u^{-1/\omega} +\zeta \bigl[ 7  \omega^2(1+3\omega)u^{1-1/\omega}  - 8  \omega^2 (1+4\omega+3\omega^2) + \omega^2 u^{-1/\omega}(1+15\omega+24\omega^2) \bigr] \\
  2 \zeta^2 \bigl[ u^{-1/\omega} \omega^2 (7+15\omega) + u (-1+\omega+12\omega^2)+(1-\omega-17\omega^2-15\omega^3)   \bigr]\Bigr)  < u.
  \end{multline}
 Again this condition  is satisfied for the acceptable region of the parameters, when we are using the quantum improved solution. For some specific values of the parameters, the comparison between classical and quantum improved solutions is presented in figure (\ref{fig4}).

\begin{figure}
    \centering
    	\includegraphics[scale=0.4]{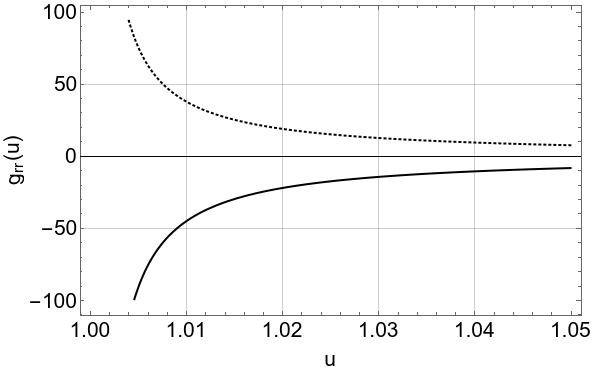}
    \caption{The radial metric component $g_{rr}$ (for spherical solution) for $(\omega,\zeta) = (-0.275,0.178)$. The dashed line is the classical solution, while the thick line is the quantum one.}
	\label{fig4}
\end{figure}
\end{itemize}

This result shows that although, there is quantum improved traversable spherical wormhole, for a wide range of the parameters $(\omega,\zeta)$, but the matter is exotic.

\section{Concluding Remarks}
We have shown that it is possible to have traversable wormholes using non--exotic matter, when one takes into account the quantum effects on gravity. The quantum effects are introduced via improvement of the field equations using the running coupling obtained from the
exact renormalization group equation to describe the
asymptotic safe gravity\cite{Reuter-1st}.

For improved pseudospherical wormholes, trip through the throat is possible using non--exotic matter unlike in the classical solution\cite{Lobo}.
To clarify this property of the improved pseudospherical wormhole, let us study the behavior of a timelike geodesic congruence on this spacetime.
For radial geodesics,  the tangent vector $u^{\alpha}$ has $ u^{\theta} = u^{\varphi} = 0  $.
Since the spacetime is static, we would have $ u^{t} = -e^{-2\Phi}$, and thus, $u^{r} = \pm \sqrt{(1-b/r) (1+e^{-2\Phi})}$. The plus sign of $u^{r}$ applies to the outgoing geodesics and the minus sign is for the ingoing ones. 

As it is usual the deviation tensor $B_{\alpha\beta} \equiv \nabla_{\beta}u_{\alpha}$, can be decomposed to trace, symmetric and antisymmetric parts, which are given as follows for our solution
\begin{align}
  & \bar{\Theta} \equiv r_t\Theta = \pm \frac{4}{u} (1-\frac{\tilde{b}}{u})^{1/2} \ , \label{expansion p.}\\
  & \bar{\sigma}_{\alpha\beta} \equiv r_t\sigma_{\alpha\beta}  =\mp(1-\frac{\tilde{b}}{u})^{1/2}
  \begin{pmatrix}
 \frac{8}{3u} & 0 & 0 & 0 \\
0 & \frac{\sqrt{2}(\dot{\tilde{b}}u-\tilde{b})}{u^2(1-\frac{\tilde{b}}{u})^{2}}+\frac{8}{3u}(1-\frac{\tilde{b}}{u})^{1/2}& 0 & 0 \\
0 & 0 & - u \frac{4-3\sqrt{2}}{3} & 0 \\
0 & 0 & 0 & - u \frac{4-3\sqrt{2}}{3}  \sinh^2\theta 
\end{pmatrix} \ ,\\
  & \bar{\omega}_{\alpha\beta} \equiv \omega_{\alpha\beta}/r_t  = 0 \ .
\end{align}
Therefore the expansion parameter, $\bar{\Theta}$, evolves along the geodesics as
\begin{equation}
  \dv{\bar{\Theta}}{\tau} = \frac{6\tilde{b}-2u(2+\dot{\tilde{b}})}{u^3 \sqrt{1-\frac{\tilde{b}}{u}}} \ .
\end{equation}
For the shape function given by equation (\ref{b-p}), this 
becomes
\[ \dv{\bar{\Theta}}{\tau} =  \frac{2 u^{-3-\frac{1}{\omega}}}{\sqrt{\omega^2(\omega-2\zeta)^2 u^{-\frac{1+\omega}{\omega}}} } \times  \{-\omega^3+\omega^4(-3+2u^{1+\frac{1}{\omega}})+  \]
\[ \zeta[\omega-2\omega^2(-9+12u^{\frac{1}{\omega}}-7u^{1+\frac{1}{\omega}})+\omega^3(69+34u^{1+\frac{1}{\omega}}-96u^{\frac{1}{\omega}})+72\omega^4(1-u^{\frac{1}{\omega}})] +
\]
\[  \zeta^2 [2u^{\frac{1}{\omega}}(-3+2u)-2\omega(7+2u^{1+\frac{1}{\omega}}-3u^{\frac{1}{\omega}})-2\omega^2(-51u^{\frac{1}{\omega}}+36+20u^{1+\frac{1}{\omega}})+90\omega^3(-1+u^{\frac{1}{\omega}})] \} \times
\]
\[  \{ \omega^4(1-u^{1+\frac{1}{\omega}})+\zeta[\omega^2(-1+8u^{\frac{1}{\omega}}-7u^{1+\frac{1}{\omega}})-\omega^3(15+17u^{1+\frac{1}{\omega}}-32u^{\frac{1}{\omega}})-24\omega^4(1-u^{\frac{1}{\omega}})] +
\]
\begin{equation} \zeta^2[2(1-u)u^{\frac{1}{\omega}}-2\omega(1-u)u^{\frac{1}{\omega}}+\omega^2(-34u^{\frac{1}{\omega}}+14+20u^{1+\frac{1}{\omega}})+30\omega^3(1-u^{\frac{1}{\omega}})]
\}^{-1/2}  \ .
\end{equation}

\begin{figure}
    \centering
		\includegraphics[width=0.50\textwidth]{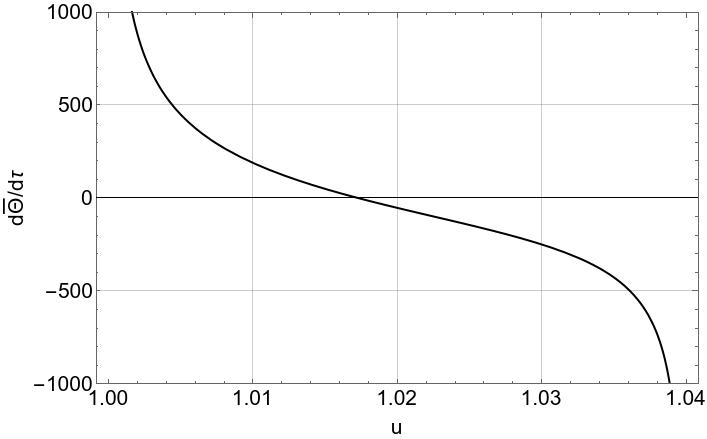}
	\caption{The variation of the expansion parameter $\bar{\Theta}$ about the $u=1$, for $(\omega,\zeta)=(0.030,0.007)$.}
	\label{fig5}
\end{figure}

This is plotted for a special value of the $(\omega,\zeta)$, satisfying the traversability conditions of section \Romannum{2}, in figure (\ref{fig5}).

As it can be seen, as the passenger approaches the throat, the rate of change of the expansion parameter, eventually becomes positive and grows up. This pushes the passenger through the throat and the wormhole becomes traversable.

\begin{figure}
    \centering
    \begin{minipage}{0.5\textwidth}  \nonumber
        \centering
        \includegraphics[width=0.9\textwidth]{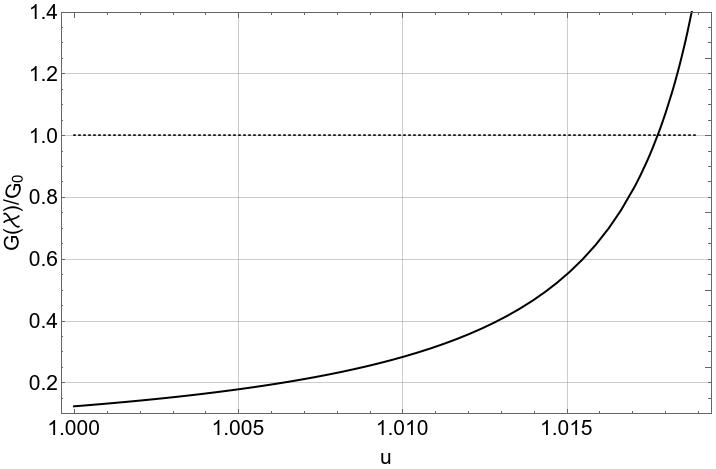}
    \end{minipage}\hfill
    \begin{minipage}{0.5\textwidth} \nonumber
        \centering
        \includegraphics[width=0.9\textwidth]{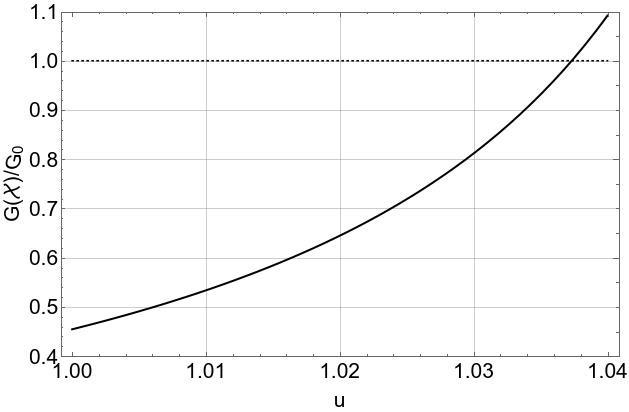} 
    \end{minipage}
    \caption{The behavior of running gravitational coupling near the throat. The left side one is for the pseudospherical wormhole solution with the acceptable parameters $(\omega,\zeta)=(0.030,0.007)$ and the right side one is for the spherical solution with $(\omega,\zeta)=(-0.200,0.200)$. As it can be seen, the quantum effects weaken the running coupling near the throat. It should be reminded that for spherical solution, exotic matter is needed.} 
    \label{fig6}
\end{figure}

This process can be clarified more by exploring the behavior of the $G(\chi)/G_0$ near the throat (see figure (\ref{fig6})). The quantum corrections weaken the classical coupling and let the matter go through the throat. This is what is expected from the antiscreening condition. After passing the throat, for the pseudospherical symmetric spacetime, it approaches to the classical one.

Another important quantity which helps us to understand the nature of such traversable wormholes is the average null energy condition (ANEC).
It can be shown that in the semiclassical approach, if the ANEC holds wormholes cannot exist\cite{Ford,Wald}.

The ANEC is weaker than the local null energy condition and requires that the integral
$\int_C T_{\alpha\beta}k^{\alpha}k^{\beta} \dd{\gamma}$, be non--negative, where $C$ is a complete null geodesic, $\gamma$ ia an affine parameter and $k^{\alpha}$ is the tangent vector along the $C$. This integral can be interpreted as the total energy flux through $C$. Thus, the satisfaction of the ANEC means that the total energy flux through any null surface must be non--negative\cite{Wald}.

For our model we have the effective (i.e. ERGE modified) energy--momentum tensor
\begin{equation}
T^{(ERGE)}_{\alpha\beta} = \frac{G(\chi)}{G_0} \left( T_{\alpha\beta}+X_{\alpha\beta}/8\pi \right)
\end{equation}
to describe the quantum corrected geometrical tensor $G_{\alpha\beta}$. Then, on the region of the validity of the null energy condition (acceptable region of the figure (\ref{fig1})) we have to check that whether
\begin{equation}
\mathcal{I}_{ANEC} \equiv \int_C T^{(ERGE)}_{\alpha\beta}k^{\alpha}k^{\beta} \dd{\gamma} > 0
\end{equation}
is satisfied for the solution \eqref{PSM} or not.
Since, using the field equations, $ T^{(ERGE)}_{\alpha\beta} \equiv G_{\alpha\beta} / 8\pi G_0$,  we would have
\begin{equation}
\mathcal{I}_{ANEC} \equiv  \frac{1}{8\pi G_0 } \int_C  G_{\alpha\beta} k^{\alpha}k^{\beta} \dd{\gamma} \ .
\end{equation}

Therefore, for our zero--tidal--force model and the null geodesic $k_{\mu}=(1,(1-b/r)^{1/2},0,0)$, the wormhole traversability depends on the negativity of 
\begin{equation}
\mathcal{I}_{ANEC} \equiv  \frac{1}{8\pi G_0 r_t} \int_C (\frac{\dot{\tilde{b}}}{u^2}-\frac{\tilde{b}}{u^3})(1-\frac{\tilde{b}}{u})^{-1/2} \dd{u} \label{ANEC}
\end{equation}
in the acceptable region of the  figure (\ref{fig1}). For any acceptable choice of the  $(\omega,\zeta)$, the negativity of this integral can be investigated by numerical methods. For example, the $(\omega,\zeta) = (0.035,0.006)$ results in $\mathcal{I}_{ANEC} = -9.397/8\pi G_0$.

\begin{figure}
    \centering
		\includegraphics[width=0.8\textwidth]{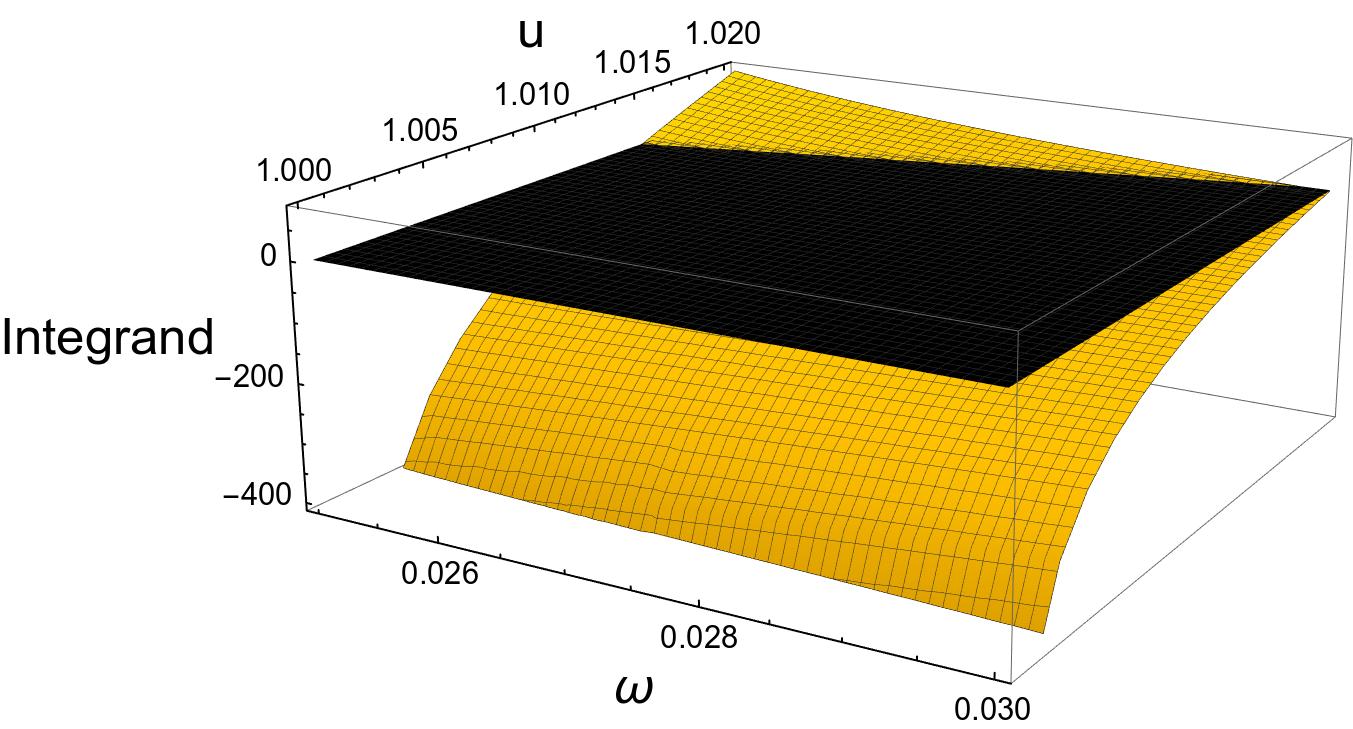}
	\caption{Integrand of ANEC for the $T^{(ERGE)}_{\alpha\beta}$ for $\zeta=0.006$, as a function of $(u,\omega)$.}
	\label{fig7}
\end{figure}

To see the general behavior, it is better to fix $\zeta$ and look at the sensitivity of the ANEC violation by varying the state equation parameter, $\omega$. We choose $\zeta=0.006$ for this purpose, and also note that the region of the validity of the solution \eqref{b-p} is just near the throat. In figure (\ref{fig7}), we have plotted the integrand of ANEC as a function of $(u,\omega)$. Since $u>1$, negative integrand is the sign of the breakdown of ANEC. As it can be seen, in the acceptable region of $\omega$ for $\zeta=0.006$ the ANEC is violated. In summary, for our solution, NEC is valid for matter energy--momentum tensor, but ANEC is violated for effective energy--momentum tensor. This second, makes the wormhole traversable, while the first is the signature of having non--exotic matter.

It should be noted at this end, that although we obtained the results for $f=\xi R$, the same results, more or less, are expected for other choices for $f$. This is because as it is shown in\cite{our4}, for any $f$, it is possible that the model shows repulsive behavior and this quantum effects can make wormholes traversable with non--exotic matter.

\vglue1cm
\textbf{Acknowledgment:} This work is supported by a grant from Iran National Science Foundation (INSF).

\appendix
\renewcommand{\thesection}{Appendix \Alph{section}:}
\section{Constraints on the physical solutions}

In this appendix, we investigate the physical constraints on the size of the psudospherical traversable solution and traveler. This is the result of setting some restrictions on the value of the tidal forces which could be felt by any physical traveler. This force is described best by the deviation equation in the curved spacetimes. It can be shown that the improved deviation equation would be\cite{our4}
\begin{equation}
  \frac{\textit{D}^2 \xi^{\alpha}}{\textit{D}\tau^2} = -R^{\alpha}_{\ \beta\gamma\sigma} u^{\beta}\xi^{\gamma}u^{\sigma} - \frac{1}{8\pi}\xi^{\beta}\nabla_{\beta}\nabla^{\gamma} \left[ \mathcal{J(\chi)}g^{\alpha\kappa}G_{\gamma\kappa} +g^{\alpha\kappa} X_{\gamma\kappa}(\chi) \right] \ ,
\end{equation}
where $\mathcal{J} \equiv G^{-1}(\chi) $ and $u^{\alpha}$ is the traveler four--velocity. 
The $\xi^{\alpha}$ is the spatial length of the traveler in its reference frame. This frame is nothing but the Lorentz transformed local frame. In the following, any quantity in this frame is distinguished by a {\em tilde} symbol like $\tilde{R}_{\alpha\beta\gamma\delta}$.

Therefore, the tidal acceleration would be
\begin{equation}
\frac{\textit{D}^2 \tilde{\xi}^{\alpha}}{\textit{D}\tau^2} = -\tilde{R}^{\alpha}_{\ \beta\gamma\sigma} \tilde{u}^{\beta}\tilde{\xi}^{\gamma}\tilde{u}^{\sigma} - \frac{1}{8\pi}\tilde{\xi}^{\beta}\tilde{\nabla}_{\beta}\tilde{\nabla}^{\gamma} \left[ \mathcal{J(\chi)}\tilde{g}^{\alpha\kappa}\tilde{G}_{\gamma\kappa} +\tilde{g}^{\alpha\kappa} \tilde{X}_{\gamma\kappa}(\chi) \right] \ .
\end{equation}
For the traveler velocity $\tilde{u}^{\alpha}=\delta^{\alpha}_{t}$ and the purely spatial length $\tilde{\xi}^{0}=0$, the $\Delta\tilde{a}^{\alpha} \equiv \textit{D}^2 \tilde{\xi}^{\alpha}/\textit{D}\tau^2$ becomes purely spatial with components\cite{Morris},
\begin{equation}
\Delta\tilde{a}^{i} = -\tilde{R}^{i}_{\ 0j0} \tilde{\xi}^{j} - \frac{1}{8\pi}\tilde{\xi}^{j}\tilde{\nabla}_{j}\tilde{\nabla}^{\gamma} \left[ \mathcal{J(\chi)}\tilde{g}^{i\kappa}\tilde{G}_{\gamma\kappa} +\tilde{g}^{i\kappa} \tilde{X}_{\gamma\kappa}(\chi) \right] \label{tidal-force} \ .
\end{equation}

In order to the traveler  can tolerate the tidal acceleration, it should have some upper bound.  Using the typical size of an astronaut and noting that the human body is adopted to earth's gravitational field such a bound can be found for astrophysical wormholes\cite{Morris}. But here, we are dealing with quantum wormholes usually of the size of planck length and present at the very early universe. For such a situation, the traveler is in fact a GUT particle and the upper bound is the energy scale of that epoch.
\begin{equation}
\abs{-\tilde{R}^{i}_{\ 0j0} \tilde{\xi}^{j} - \frac{1}{8\pi}\tilde{\xi}^{j}\tilde{\nabla}_{j}\tilde{\nabla}^{\gamma} \left[ \mathcal{J(\chi)}\tilde{g}^{i\kappa}\tilde{G}_{\gamma\kappa} +\tilde{g}^{i\kappa} \tilde{X}_{\gamma\kappa}(\chi) \right]} < \mathcal{S}_c \ , 
\end{equation}
where $\mathcal{S}_c$ is the scaling energy at each epoch. We use this to estimate the maximum size of the traveler through the wormhole solution \eqref{PSM}, before collapsing.

By Lorentz transforming the local Riemann curvature of the pseudospherical solution \eqref{PSM}, we would have
\begin{align}
 \tilde{R}_{rtrt} = \bar{R}_{rtrt} &= -(1-\frac{b}{r}) \left(-\Phi''+\frac{b'r-b}{2r(r-b)}\Phi'-\Phi'^2 \right) \ , \\
 \tilde{R}_{\theta t\theta t} = \tilde{R}_{\phi t\phi t}  &= \gamma^2 \bar{R}_{\theta t \theta t} + \gamma^2 v^2 \bar{R}_{\theta r \theta r} \nonumber \\
& = \frac{\gamma^2}{2r^2} \left[ v^2 (b'-\frac{b}{r}) +2r(1-\frac{b}{r})\Phi' \right] 
\end{align}
where $\gamma=(1-v^2)^{-1/2}$.

Thus, for zero--tidal--force solution $(\Phi=0)$ the equation \eqref{tidal-force} would be
\begin{align}
& \Delta\tilde{a}^{r} = - \frac{1}{8\pi}\tilde{\xi}^{j}\tilde{\nabla}_{j}\tilde{\nabla}^{\gamma} \left[ \mathcal{J(\chi)}\tilde{g}^{r\kappa}\tilde{G}_{\gamma\kappa} +\tilde{g}^{r\kappa} \tilde{X}_{\gamma\kappa}(\chi) \right] \ , \label{RTA} \\
& \Delta\tilde{a}^{\theta} =  -\tilde{\xi}^{\theta}  \frac{\gamma^2}{2r^2} \left[ v^2 (b'-\frac{b}{r}) \right] \label{LTA} \ .
\end{align}
and thus
\begin{align}
& \abs{- \frac{1}{8\pi}\tilde{\xi}^{j}\tilde{\nabla}_{j}\tilde{\nabla}^{\gamma} \left[ \mathcal{J(\chi)}\tilde{g}^{r\kappa}\tilde{G}_{\gamma\kappa} +\tilde{g}^{r\kappa} \tilde{X}_{\gamma\kappa}(\chi) \right]} < \mathcal{S}_c \ , \label{1st-Con} \\
& \abs{- \tilde{\xi}^{\theta}  \frac{\gamma^2}{2r^2} \left[ v^2 (b'-\frac{b}{r}) \right]} < \mathcal{S}_c \ . \label{2nd-Con}
\end{align}

As stated before, to have a non--exotic wormhole solution, the values $(\omega,\zeta)$ are restricted to specific domains (see figure (\ref{fig1})). For $(\omega,\zeta)=(0.035,0.006)$ which is a suitable choice, and at the energy scale of the very early universe, $\mathcal{S}_c \simeq 10^{19}$(GeV), the throat radius becomes $r_t\simeq 14 \ell_P$. Simple calculations shows that equation (\ref{1st-Con}) restricts the size to $\abs{\xi^{r}}\simeq 1.317\ell_P$ and assuming  $\abs{\xi^{\theta}} \sim \abs{\xi^{r}}$, equation \eqref{2nd-Con} gives $v < 0.722(c)$. As a result a traveler of size of order planck length with ultra-relativistic speed can have a trip through this quantum wormhole.

\end{document}